\documentclass[10pt]{article}
\usepackage{epsfig,amsmath,amsthm,graphicx,amssymb,overpic}
\usepackage{epsfig,amsmath,graphicx,amssymb,overpic,cite}
\usepackage{listings,diagbox}

\def\be{\begin{equation}}
\def\ee{\end{equation}}
\def\bee{\begin{eqnarray}}
\def\ene{\end{eqnarray}}
\def\bes{\begin{subequations}}
\def\ees{\end{subequations}}

\def\v{\vspace{0.1in}}

\newcommand{\PT}{\mathcal{PT}}

\begin{document}

\baselineskip=12pt
\renewcommand {\thefootnote}{\dag}
\renewcommand {\thefootnote}{\ddag}
\renewcommand {\thefootnote}{ }

\pagestyle{plain}

\begin{center}
\baselineskip=16pt \leftline{} \vspace{-.3in} {\Large \bf Solving forward and inverse problems of the logarithmic nonlinear Schr\"odinger equation with $\PT$-symmetric harmonic potential via deep learning} \\[0.2in]
\end{center}

\begin{center}
Zijian Zhou and Zhenya Yan$^{*}$\footnote{$^{*}${\it Email address}: zyyan@mmrc.iss.ac.cn (Corresponding author)}  \\[0.03in]
{\it Key Laboratory of Mathematics Mechanization, Academy of Mathematics and Systems Science, \\ Chinese Academy of Sciences, Beijing 100190, China \\
 School of Mathematical Sciences, University of Chinese Academy of Sciences, Beijing 100049, China} \\
\end{center}


{\baselineskip=12pt


\vspace{0.18in}

\noindent {\bf Abstract:}\, In this paper, we investigate the logarithmic nonlinear Schr\"odinger (LNLS) equation with the parity-time ($\PT$)-symmetric harmonic potential, which is an important physical model in many fields such as nuclear physics, quantum optics, magma transport phenomena, and effective quantum gravity. Three types of initial value conditions and periodic boundary conditions are chosen to solve the LNLS equation with $\PT$-symmetric harmonic potential via the physics-informed neural networks (PINNs) deep learning method, and these obtained results are compared with ones deduced from the Fourier spectral method. Moreover, we also investigate the effectiveness of the PINNs deep learning for the LNLS equation with $\PT$ symmetric potential by choosing the distinct space widths or distinct optimized steps. Finally, we use the PINNs deep learning method to effectively tackle the data-driven discovery of the LNLS equation with $\PT$-symmetric harmonic potential such that the coefficients of dispersion and nonlinear terms or the amplitudes of $\PT$-symmetric harmonic potential can be approximately found.


\vspace{0.1in} \noindent {\it Keywords:} Logarithmic nonlinear Schr\"odinger equation; $\PT$-symmetric potentials; physics-informed neural networks; solitons



\centerline{\it Phys. Lett.  A  387 (2021) 127010.}

\section{Introduction}

With the rapid development of modern science and high technology, more and more massive data sets are generated in many scientific fields and real lives. It is still a huge and important issue that how well to mine and further powerfully apply them in the scientific fields and even our lives.
Though the artificial intelligence (AI) has experienced several stages of development, but now it seems that the new generation of AI is facing a new  challenge or revolution, and changing our lives. The deep (machine) learning based on the neural networks (NNs), as one of most significant approaches realizing AI, has been applied in various of fields, such as image recognition, computer vision, natural language processing, solving equations, and other business fields~\cite{dl, dl2}. Recently, based on the neural networks and automatic differentiation (AD)~\cite{ad1,ad2,ad3}, some deep learning approaches had been presented to approximate the solutions of some types of differential equations (see, e.g.,
Refs.~\cite{nn1,nn2,nn3,siri2018,raissi2018,raissi2019,pang2019,yang2018,zhang2019,zhang2019b,nabian2018,chri2019,lu2019,mich2020} and references therein), in which the physics-information neural networks (PINNs) deep learning method~\cite{raissi2019} is a simple and effective method to solving the partial differential equations (PDEs) via deep learning, and was extended to other types of differential equations such as integro-differential equations, fractional PDEs, and stochastic PDEs~\cite{pang2019,yang2018,zhang2019,zhang2019b,nabian2018,chri2019,lu2019}. More recently,
a Python library DeepXDE was presented for some PINNs approaches~\cite{lu2019}. The same characteristic of these deep neural network methods is
to use the NNs instead of the traditional numerical discretization (ND) methods~\cite{lu2019}. Moreover, the designs of these NNs seem to be easily carried out than the traditional ND methods.


 The PINNs deep learning method~\cite{raissi2019,lu2019} can be used to study not only data-driven solutions of some types of differential equations, but data-driven discovery of PDEs. Some other methods were also presented for the data-driven discovery of PDEs~\cite{rudy2017,raissi2017,raissi2017b}. In particular, this PINNs deep learning approach~\cite{raissi2019,lu2019} can be used to study the nonlinear wave equations such as the physically interesting soliton equations. In the following, we simply introduce the PINNs method solving data-driven solutions of the PDEs  with initial-boundary value conditions~\cite{raissi2019}. For the given nonlinear wave equation with the initial-boundary value (IBV) conditions
\bee\label{pde}
\left\{\begin{array}{ll}
iu_t={\mathcal L}{\mathcal N}[x, u, u^*; \Lambda_{0}]),\quad & x\in D\setminus \partial D\subset \mathbb{R}^m, \,\, t\in (0, T],\v\\
I[u(x,t)]_{t=0}=u_{I}(x), \quad & x\in D \,\, \text{(initial condition)}, \v \\
B[u(x, t)]|_{x\in \partial D}=u_{B}(t),\quad & t\in [0, T],\,\, \text{(boundary conditions)},
\end{array}\right.
\ene
where $u=u(x, t)$ stands for the latent (hidden) solution, ${\mathcal L}{\mathcal N}$ is the combination of  linear and nonlinear operators
parametrized by the vector $\Lambda_0$, $I[\cdot],\, B[\cdot]$ are the operators, $I[u(x,t)]_{t=0}=u_{I}(x)$ and $B[u(x,t)]|_{x\in \partial D}=u_{B}(t)$ stand for the initial conditions, and the Dirichlet (or Neumann, or Robin) boundary conditions, respectively.
Let a deep hidden physics model $F(x,t)$ be
\bee\label{F}
F(x,t):=iu_t-{\mathcal L}{\mathcal N}[x, u, u^*; \Lambda_{0}]).
\ene
Then one can obtain the derivatives of the NN $u$ with respect to time $t$ and space $x$ by applying the chain rule for differentiating compositions of functions using automatic differentiation~\cite{ad1,ad2,ad3}. It is worth emphasizing that automatic differentiation is different from, and in several aspects superior to, numerical or symbolic differentiation; two commonly encountered techniques of computing derivatives~\cite{ad1,ad2,ad3}. To perform the derivatives in the above-mentioned definition one can depend on the Tensorflow~\cite{tensorflow}, which is a popular and relatively well documented open source software library for automatic differentiation and deep learning computations. The PINNs approach exhibits the structure of the NN, which has the $i$ hidden Layers, and every hidden layer has $n$ neurons (see Fig.~\ref{fig1-nn}). The nonlinear activation function in the NN is chosen as follows~\cite{raissi2019}
\begin{equation}
\label{tanh}
    A_j=\tanh(w_j*A_{j-1}+b_j),
\end{equation}
where the $w_j$ is a ${\rm dim}(A_j)\times {\rm dim}(A_{j-1})$ matrix, $A_0,A_{i+1},b_{i+1}\in \textbf{R}^2$, $A_j,b_j\in\textbf{R}^n$. The training loss function is given by
\bee\label{tl}
{\rm TL}=\frac{1}{N_I}\sum_{j=1}^{N_I}|I[u(x_I^j,t)]_{t=0}-u_{I}(x_I^j)|^2+\frac{1}{N_B}\sum_{j=1}^{N_B}|B[u(x_B, t_B^j)]|_{x_B\in \partial D}-u_{B}(t_B^j)|^2 +\frac{1}{N_S}\sum_{j=1}^{N_S}|F(x_S^j, t_S^j)|^2,
\ene
where $\{x_I^j, u_I^j\}_{j=1}^{N_I}$ and $\{t_B^j, u_B^j\}_{j=1}^{N_B}$ denote the initial and boundary training data on $u(x,t)$, respectively,
and $\{x_S^j, t_S^j,\, F(x_S^j, t_S^j)\}_{j=1}^{N_S}$ specify the collocations points for $F(x, t)$. The optimization method for all loss functions
was chosen as the L-BFGS algorithm~\cite{liu89}.

\begin{figure}[!t]
\begin{center}
\vspace{0.05in} 
{\scalebox{0.4}[0.4]{\includegraphics{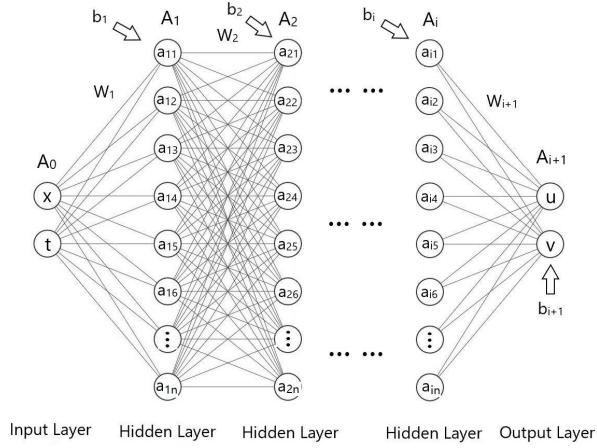}}}
\end{center}
\par
\vspace{-0.1in}
\caption{\protect\small The deep learning frame based on the PINNs.}
\label{fig1-nn}
\end{figure}

The main steps of the PINNs deep learning method solving the PDE (\ref{pde}) with IBV conditions (see Fig.~\ref{fig1-nn}) are summarized as follows~\cite{raissi2019,lu2019}:
\begin{itemize}
\item Step 1.  Introduce a neural network $u_{\rm NN}(x,t; {\bf w}, {\bf b})$ with parameters ${\bf w}=\{w_j\}_{1}^{i}$
 and ${\bf b}=\{b_j\}_{1}^{i}$ being the weight matrices and bias vectors, and the PINN $F(x,t)$ given by Eq.~(\ref{F}) in terms of the back-propagating AD~\cite{back};

\item Step 2:\, Extract the three training sets ${\rm TS}_I,\, {\rm TS}_B,\, {\rm TS}_E$ for the initial value condition, boundary value conditions, and considered model;

\item Step 3:\, Extract a training loss function ${\rm TL}$ given by Eq.~(\ref{tl}) by summing the weighted $\mathbb{L}^2$-norm error of both the $F(x,t)$ and IBV residuals;

\item Step 4:\, Train the NN to find the best parameters $\{\widehat{\bf w}, \, \widehat{\bf b}\}$ by minimizing the loss function ${\rm TL}$ in terms of the L-BFGS optimization algorithm~\cite{liu89}.

\end{itemize}

Since the concept of the $\PT$ symmetry was introduced in 1998~\cite{pt1}, the $\PT$ symmetry plays an more and more important role in the generalized Hamiltonian systems and nonlinear wave equations (see, e.g., Refs.~\cite{pt2,pt3,pt4,pt5,pt6,pt7,pt8,pt9,pt10,pt11,pt12,pt13,pt14,pt15,pt16,pt17,pt18,boris15, chen17, yan16,yan17, chen18, abdu18,miha12} and references therein). It is still a significant subject to solve the nonlinear wave equations with $\PT$-symmetric potentials~\cite{pt12}. To the best of our knowledge, the deep learning method was not used to study the nonlinear wave equations with $\PT$-symmetric potentials yet. In this paper we would like to study the solutions of the logarithmic nonlinear Schr\"odinger (LNSL) equation with $\PT$-symmetric potentials via the PINNs deep learning method.

The rest of this paper is arranged as follows. In Sec. 2, we firstly introduce the deep learning scheme for the LNLS equation with $\PT$-symmetric potentials. And then we use the effect deep learning scheme to investigate the LNLS equation with $\PT$-symmetric potentials under the distinct initial-boundary value conditions. In Sec. 3, we investigate the effectiveness of the PINNs deep learning method in the LNLS equation (\ref{lnls}) with IBV conditions by comparing distinct space widths or  optimized steps. In Sec. 4, we use the PINNs deep learning method to tackle the data-driven discovery of the LNLS equation with $\PT$-symmetric harmonic potentials. Finally, we give some conclusions and discussions.

\section{Data-deriven solutions of the LNLS equation with $\PT$-symmetric potential}

\subsection{The scheme for the PINNs deep learning method}

The well-known nonlinear Schr\"odinger (NLS) equation is a fundamental model, and of important significance in some physical fields such as nonlinear optics, Bose-Einstein condensates, deep ocean, DNA, even finance.  In 1975, Bialynicki-Birula and Mycielski~\cite{lse1} first presented an extended NLS equation, that is, the NLS equation with a logarithmic nonlinear term $g\ln(a|\varphi|^2)\varphi$ (simply called the LNLS equation)
\begin{equation}\label{lnls}
    i\hbar\varphi_t=-\frac{\hbar^2}{2m}\varphi_{xx}+V(x)\varphi+\sigma\ln(a|\varphi|^2)\varphi,
\end{equation}
where $\varphi=\varphi(x,t)$ stands for the complex field, $\hbar$ is the reduced Planck constant, $m$ is the mass, $\sigma$ is the nonlinear coefficient corresponding to the focusing $(\sigma<0)$ and defocusing $(\sigma>0)$ interactions, $V(x)$ usually denotes the real-valued potential, $a$ is a positive, and real constant and can be chosen as $a=1$, without loss of generality, by adding a constant to the potential~\cite{lse2}.
Eq.~(\ref{lnls}) can be applied to quantum mechanics~\cite{lse1,lse2,lse3},  dissipative systems~\cite{ds},  nuclear physics~\cite{np,np2}, quantum optics~\cite{qo},
magma transport phenomena~\cite{tdp}, open quantum systems, effective quantum gravity, theory of superfluidity and Bose-Einstein
condensation~\cite{other1, other2, other3,other4,shuai19}.

We here use the above-mentioned PINNs deep learning method~\cite{raissi2019} to study the data-driven solutions of the dimensionless LNLS equation with the complex $\PT$-symmetric potential
\bee\label{lnls}
 i\varphi_t=-\varphi_{xx}+[V(x)+iW(x)]\varphi+\sigma\ln(|\varphi|^2)\varphi,\quad x\in (-L, L),\quad t\in (0, T),
\ene
with the following initial-boundary value conditions
\bee\label{lnls-ivp}
\begin{array}{l}
 \varphi(x,0)=\varphi_0(x),\quad  x\in [-L, L], \v\\
 \varphi(-L, t)=\varphi(L, t),\quad  t\in [0, T],
\end{array}
\ene
where $\varphi=\varphi(x,t)$ is a complex field, and $V(x)$ and $W(x)$ are real-valued functions of space, and denote the external potential and the gain-and-loss distribution, respectively, The required conditions $V(x)=V(-x)$ and $W(x)=-W(-x)$ imply that the complex potential $ V(x)+iW(x)$ is $\PT$-symmetric~\cite{pt1}, and are chosen as the physically interesting potentials (also called the
$\PT$-symmetric harmonic potential)~\cite{chen18}
\begin{eqnarray}\label{potential-A}
V(x)=V_0 x^2,\qquad W(x)=W_0 x,
\end{eqnarray}
where $V_0>0,\, W_0$ are real-valued constants, and used to modulate the effect of $\PT$-symmetric potential, $V(x)$ is the well-known harmonic potential, and the gain-and-loss distribution $W(x)$ is not limited (except for $W_0=0)$, and always has the effect on the LNLS equation (\ref{lnls}). The constant $\sigma$ stands for the nonlinear coefficients, and can be chosen as $\sigma=\pm 1$ without loss of generality. The cases $\sigma=-1$ and $\sigma=1$ correspond to the self-focusing and defocusing interactions, respectively.

Since the function $\varphi(x,t)$ in Eq.~(\ref{lnls}) is complex, thus it can be rewritten as $\varphi(x,t)=u(x,t)+iv(x,t)$ with $u(x,t),\,v(x,t)$ being its real and imaginary parts, respectively. Introduce the associated complex-valued PINNs $F(x,t)=iF_u(x,t)-F_v(x,t)$ with $-F_v(x,t),\,F_u(x,t)$ being its real and imaginary parts, respectively, as
\bee
 \begin{array}{l}
 F(x,t):=i\varphi_t+\varphi_{xx}-[V(x)+iW(x)]\varphi-\sigma \varphi\ln(|\varphi|^2), \v\\
 F_u(x,t):= u_t + v_{xx} - W(x)u - V(x)v - \sigma v\ln(u^2 + v^2), \v\\
 F_v(x,t):= v_t - u_{xx} + V(x)u - W(x)v + \sigma u\ln(u^2 + v^2),
 \end{array}
  \ene
and proceed by approximating $\varphi(x,t)$ by a complex-valued deep neural network.  The complex neural network $\varphi(x,t)=(u(x,t), v(x,t))$ can be defined as
\begin{lstlisting}
def psi(x, t):
    psi = neural_net(tf.concat([x,t],1), weights, biases)
    u = psi[:,0:1]
    v = psi[:,1:2]
    return u, v
\end{lstlisting}

Moreover, the physics-informed neural network $F(x,t)$ is chosen the form
\begin{lstlisting}
 def F(x, t):
     u, v = psi(x, t)
     u_t = tf.gradients(u, t)[0]
     u_x = tf.gradients(u, x)[0]
     u_xx = tf.gradients(u_x, x)[0]
     v_t = tf.gradients(v, t)[0]
     v_x = tf.gradients(v, x)[0]
     v_xx = tf.gradients(v_x, x)[0]
     F_u = u_t + v_xx - W*u - V*v - sigma*tf.ln(u**2 + v**2)*v
     F_v = v_t - u_xx + V*u - W*v + sigma*tf.ln(u**2 + v**2)*u
     return F_u, F_v
\end{lstlisting}

The shared parameters between the neural network $\varphi(x,t)=u(x,t)+iv(x,t)$ and $F(x,t)=iF_u(x,t)-F_v(x,t)$ can be trained by minimizing the training loss (TL), that is, the sum of the TLes on the initial data (${\rm TL}_{I}$), boundary data (${\rm TL}_{B}$), and the whole equation (${\rm TL}_{S}$)
 \bee
  {\rm TL}={\rm TL}_{I}+{\rm TL}_{B}+{\rm TL}_{S},
  \ene
where the mean squared errors are chosen as
\bee\begin{array}{l}
 {\rm TL}_{I}=\displaystyle\frac{1}{N_I}\sum_{j=1}^{N_I}\left(\left|u(x_{I}^j, 0)-u_{0}^j\right|^2+\left|v(x_{I}^j, 0)-v_{0}^j\right|^2\right),\v\\
 {\rm TL}_{B}=\displaystyle\frac{1}{N_B}\sum_{j=1}^{N_B}\left(\left|u(-L, t_{B}^j)-u(L, t_{B}^j)\right|^2+\left|v(-L, t_{B}^j)-v(L, t_{B}^j)\right|^2\right),\v\\
 {\rm TL}_{S}=\displaystyle\frac{1}{N_S}\sum_{j=1}^{N_S}\left(\left|F_u(x_S^j, t_S^j)\right|^2+\left|F_v(x_S^j, t_S^j)\right|^2\right)
\end{array}
\ene
with $\{x_I^j,\, u_{0}^j,\, v_{0}^j\}_{j=1}^{N_I}$ standing for the initial data ($\varphi_0(x)=u_0(x)+iv_0(x)$),  $\{t_B^j,\, u(\pm L, t_{B}^j),\, v(\pm L, t_{B}^j)\}_{j=1}^{N_B}$ standing for the boundary data,  $\{x_S^j,\, t_S^j,\,F_u(x_S^j, t_S^j),\, F_v(x_S^j, t_S^j)\}_{j=1}^{N_S}$ representing the collocation points within a spatio-temporal region on $F(x,t)=iF_u-F_v$. In other words, $ {\rm TL}_{I}$ represents the $\mathbb{L}^2$-norm error in initial points, ${\rm TL}_{B}$ denotes the $\mathbb{L}^2$-norm error in the periodic boundary condition, and ${\rm TL}_{S}$ corresponds the $\mathbb{L}^2$-norm loss in the interior of the spatio-temporal region. All of these sampling points can be generated using a space filling Latin Hypercube Sampling strategy~\cite{stein87}.

To solve Eq.~(\ref{lnls}) with the above-mentioned initial-boundary value conditions via the PINNs  deep learning method~\cite{raissi2019}, we need to generate a LNLS.mat file by utilizing the Fourier spectral method to Eq.~(\ref{lnls}) with the initial-boundary conditions~(\ref{lnls-ivp}) in Matlab. In fact, we here choose the Fourier spectral method~\cite{yang2010} with a special Fourier discretization with 256 space modes and a fourth-order explicit Runge-Kutta temporal integrator with a time step length  $\Delta t=10^{-4}$ to simulate it. To further study Eq.~(\ref{lnls}) via the deep learning,
we choose a $201$ temporal sampling points with the same time interval such that $\varphi(x,t)=u(x,t)+iv(x,t)$ is a $256\times 201$ matrix. All of these sampling points will be packaged as a LNLS.mat file.

\subsection{Applications in the distinct initial value conditions and same boundary value condition}

In the follows, we would like to use the above-mentioned PINNs deep learning approach to investigate the solitons of the LNLS equation with $\PT$-symmetric harmonic potential (\ref{lnls}) with distinct initial-boundary value problems (\ref{lnls-ivp}).

\subsubsection{The initial value condition  arising from the exact soliton}

 We can find that Eq.~(\ref{lnls}) with the $\PT$-symmetric harmonic potential (\ref{potential-A}) admits the exact bright soliton
\bee\label{solu}
    \varphi(x,t)=A\exp\!\left(-\omega x^2\right)e^{-i[W_0x/(4\omega)+\mu t]},
\ene
with $A\in\mathbb{R},\, A\not=0$ and
\bee\label{para}
\omega=\frac14\left(\sqrt{\sigma^2+4V_0}-\sigma\right)>0,\quad \mu=2\omega\!+\!\sigma\ln A^2\!+\!\frac{W_0^2}{16\omega}.
\ene
When $|x|\to\infty$, $|\varphi(x,t)|\to 0$ and $\int_{-\infty}^{\infty}|\varphi(x,t)|^2dx=A^2\sqrt{\frac{\pi}{2\omega}}$.\\

{\it Case 1.} The focusing case $\sigma=-1$ and other parameters $W_0=0.2,\, V_0=1,\, A=1$.

We first use the exact solution (\ref{solu}) to choose the initial conditions: $\varphi_{0}(x)=\varphi(x,0)=u_{0}(x)+iv_{0}(x)$ with
\bee
u_{0}(x)=e^{-\omega x^2}\!\cos(0.2x/(4\omega)),\quad
v_{0}(x)=-e^{-\omega x^2}\!\sin(0.2x/(4\omega)),
\ene
with $\omega=(\sqrt{5}+1)/4$
and the periodic boundary condition $\varphi(-L,t)=\varphi(L, t)$ such that we can generate a high-resolution data set by using the Fourier spectral method to simulate Eq.~(\ref{lnls}). Under the data-derived setting, all measurements we observe are $\{x_I^j,\, u_{0}^j,\, v_{0}^j\}_{j=1}^{N_I}$ of the latent function $\varphi(x,t)=u(x,t)+iv(x,t)$ at initial time $t=0$ with the periodic boundary conditions $\{t_B^j,\, u(\pm L, t_{B}^j),\, v(\pm L, t_{B}^j)\}_{j=1}^{N_B}$. The training set consists of a total of $N_I=100$
data points on the initial conditions $\{u_{0}(x),\, v_{0}(x)\}$ randomly deduced from the full high-resolution data set, and $N_B=200$ randomly sampled collocation points on the periodic boundary conditions $\{u(\pm L, t),\, v(\pm L, t)\}$. We here choose $N_s=10,000$ randomly sampled collocation points used to study the LNLS Eq.~(\ref{lnls}) inside the solution region. Notice that all randomly sampled point locations are obtained via the Latin Hypercube Sampling strategy~\cite{stein87}.

We choose an $11-$layer deep neural network with $50$ neurons per layer and a hyperbolic tangent activation function given by Eq.~(\ref{tanh})
to jointly represent the latent function $\varphi(x,t)=u(x,t)+iv(x,t)$. We choose the space interval as $[-5,5]$ (i.e., $L=5$),
and the time interval as $[0, 5]$ (i.e., $T=5$). As a consequence, Figs.~\ref{fig2-exact}(a1-d1) illustrate the results of the simulations.
Fig.~\ref{fig2-exact}(a1) displays the magnitude of the predicted solution $|\varphi(x,t)|=\sqrt{u^2(x,t)+v^2(x,t)}$ for the given
locations of the initial-boundary conditions.  The whole training time of this result is $18475$ seconds on a matebook x pro with a 1.6GHz four-cores i5 processor (the same hereinafter). The relative $\mathbb{L}^2-$norm error
of $\varphi(x,t)$ in this case is $1.1791\cdot 10^{-3}$. The relative $\mathbb{L}^2-$norm errors  of $u(x,t)$ and $v(x,t)$, respectively, are $1.4040\cdot 10^{-3}$ and $1.3955\cdot 10^{-3}$. Figs.~\ref{fig2-exact}(b1-d1) exhibit the comparisons between exact and predicted solutions for distinct time instants $t=1.25,\, 2.5,\, 3.75$.\\

{\it Case 2.} The defocusing case $\sigma=1$ and other parameters $W_0=0.1, V_0=A=2$.

Similarly to Case 1, we  use the exact solution (\ref{solu}) to choose the initial conditions: $\varphi_{0}(x)=\varphi(x,0)=u_{0}(x)+iv_{0}(x)$ with
\bee
u_{0}(x)=2e^{-0.5 x^2}\!\cos(0.05x),\quad
v_{0}(x)=-2e^{-0.5 x^2}\!\sin(0.05x),
\ene
and the periodic boundary condition $\varphi(-L,t)=\varphi(L, t)$ such that we can generate a high-resolution data set by using the Fourier spectral method to simulate Eq.~(\ref{lnls}).

 The training set is composed of a total of $N_I=100$ data points on the initial conditions $\{u_{0}(x),\, v_{0}(x)\}$ randomly deduced from the full high-resolution data set, and $N_B=200$ randomly sampled collocation points on the periodic boundary conditions $\{u(\pm L, t),\, v(\pm L, t)\}$. We here choose $N_s=10,000$ randomly sampled collocation points used to study the LNLS Eq.~(\ref{lnls}) inside the solution region. Notice that all randomly sampled point locations were obtained via the Latin Hypercube Sampling strategy~\cite{stein87}.

Similarly to Case 1, we still choose an $11$-layer deep neural network with $50$ neurons per layer and a hyperbolic tangent activation function given by Eq.~(\ref{tanh}) to jointly represent the latent function $\varphi(x,t)=u(x,t)+iv(x,t)$. We take the space interval as $[-5,5]$ (i.e., $L=5$),
and the time interval as $[0, 5]$ (i.e., $T=5$). As a consequence, Figs.~\ref{fig2-exact}(a2-d2) illustrate the results of the simulations.
Fig.~\ref{fig2-exact}(a2) displays the magnitude of the predicted solution $|\varphi(x,t)|=\sqrt{u^2(x,t)+v^2(x,t)}$ for the given
locations of the initial-boundary conditions.  The whole training time of this result is $42417$ seconds. 
The relative $\mathbb{L}^2-$norm error of $\varphi(x,t)$ in this case is $1.7889\cdot 10^{-4}$. The relative $\mathbb{L}^2-$norm errors  of $u(x,t)$ and $v(x,t)$, respectively, are $8.3298\cdot 10^{-3}$ and $6.8020\cdot 10^{-3}$. Figs.~\ref{fig2-exact}(b2-d2) exhibit the comparisons between exact and predicted solutions for distinct time instants $t=1.25,\, 2.5,\, 3.75$.
\begin{figure}[!t]
\begin{center}
\vspace{0.05in} 
{\scalebox{0.56}[0.56]{\includegraphics{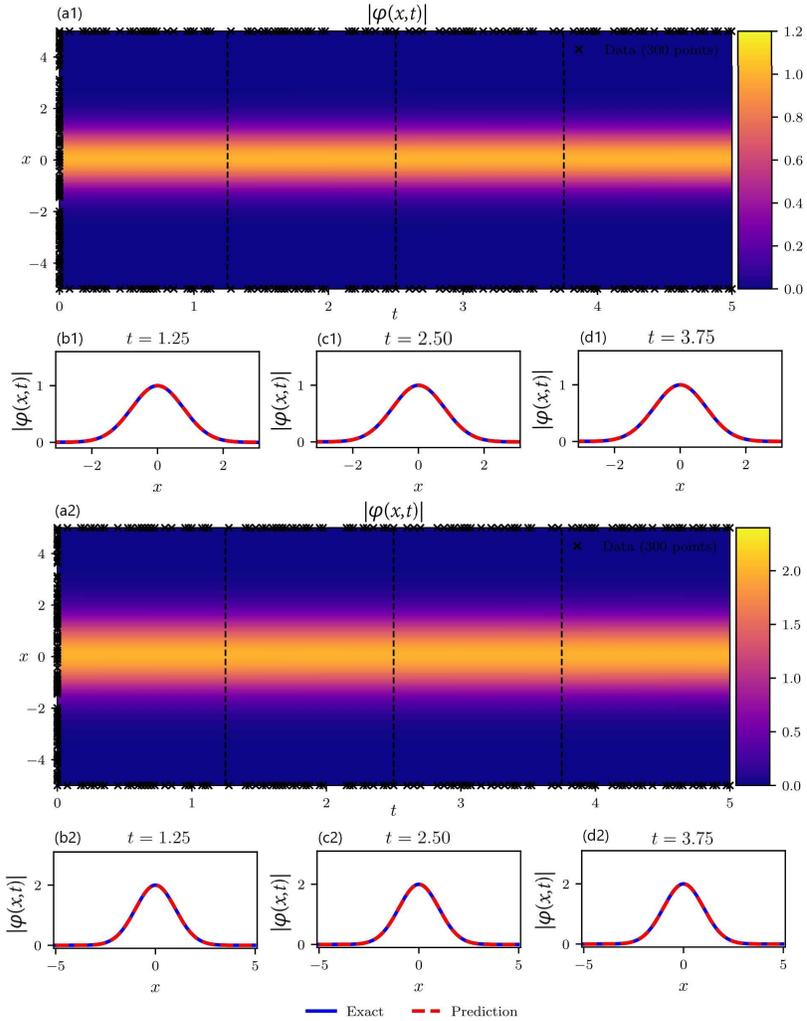}}}
\end{center}
\par
\vspace{-0.25in}
\caption{\protect\small The LNLS equation with $\PT$-symmetric potential and exact solution (\ref{solu}):
(a1) $\sigma=-1$ The predicted solution $|\varphi(x,t)|$ with $N_I=100$ data points on initial conditions and $N_B=200$ data points on periodic boundary conditions (see the cross), and $N_S=10,000$ collocation points deduced by Latin Hypercube Sampling strategy. The relative $\mathbb{L}^2-$norm error of $\varphi(x,t)$ in this case is $1.1791\cdot 10^{-3}$. (b1)-(d1) $\sigma=-1$ and (b2)-(d2) $\sigma=1$: Comparisons between the exact and predicted solutions by neural network at $t=1.25, t=2.50, t=3.75$. (a2) $\sigma=1$ The predicted solution $|\varphi(x,t)|$ with $N_I=100$ data points on initial conditions, $N_B=200$ data points on periodic boundary conditions (see the cross), and $10,000$ collocation points deduced by Latin Hypercube Sampling strategy. The relative $\mathbb{L}^2-$norm error of $\varphi(x,t)$ in this case is $1.7889\cdot 10^{-3}$. (b2)-(d2) $\sigma=1$: Comparisons between the exact and predicted solutions by neural network at $t=1.25, t=2.50, t=3.75$. }
\label{fig2-exact}
\end{figure}

\subsubsection{The initial condition arising from the non-stationary solutions}

In the second example, we consider Eq.~(\ref{lnls}) with another initial condition (which does not satisfy the stationary equation of Eq.~(\ref{lnls}))
\begin{eqnarray}\label{aexact}
   \varphi_0(x)=A\exp\left(-\omega x^2\right)e^{i\frac{W_0}{4\omega}x},
\end{eqnarray}
with $\omega$ given by Eq.~(\ref{para}), the periodic boundary conditions $\varphi(-3, t)=\varphi(3,t)$, and parameters  $W_0=0.2$, $V_0=1$, $A=2$, $\sigma=-1$ such that we have $\varphi_{0}(x)=\varphi(x,0)=u_{0}(x)+iv_{0}(x)$ with
\bee
u_{0}(x)=2e^{-\omega x^2}\!\cos(0.2x/(4\omega)),\quad v_{0}(x)=2e^{-\omega x^2}\!\sin(0.2x/(4\omega)),
\ene
with $\omega=(\sqrt{5}+1)/4$.

To solve Eq.~(\ref{lnls}) with the above-mentioned initial-boundary value conditions via the deep learning, we firstly obtain a high-resolution data set by using the above-mentioned Fourier spectral method to simulate Eq.~(\ref{lnls}) with the initial condition (\ref{aexact}) and periodic boundary conditions. As a result, we find that the amplitude of vibration of solitary wave grows as $|W_0|$ of gain-and-loss distribution increases. Generally speaking, we can
 select a 200-dimensional temporal sampling points, which may be of the equal interval. All of these sampling points will be packaged as a .mat file.

We now choose a $5$-layer deep neural network  with $90$ neurons per layer and a hyperbolic tangent activation function (\ref{tanh}) to jointly represent the latent function $\varphi(x,t)=u(x,t)+iv(x,t)$ where $u(x,t),\, v(x,t)$ represent the real and imaginary parts of $\phi(x,t)$, respectively. Fig.~\ref{fig3-aexact}a displays the magnitude of the predicted solution $|\varphi(x,t)|=\sqrt{u^2(x,t)+v^2(x,t)}$ for the given
locations of the initial-boundary conditions.  The whole training time of this result is $46397$ seconds. The relative $\mathbb{L}^2-$norm error
of $\varphi(x,t)$ in this case is $4.6204\cdot 10^{-3}$. The relative $\mathbb{L}^2-$norm errors  of $u(x,t)$ and $v(x,t)$, respectively, are $8.4528\cdot 10^{-3}$ and $8.6211\cdot 10^{-3}$. Figs.~\ref{fig3-aexact}(b-d) exhibit the comparisons between exact and predicted solutions for distinct time instants $t=2.5,\, 5.0,\, 7.5$.

\begin{figure}[!t]
\begin{center}
\vspace{0.05in} 
{\scalebox{0.56}[0.56]{\includegraphics{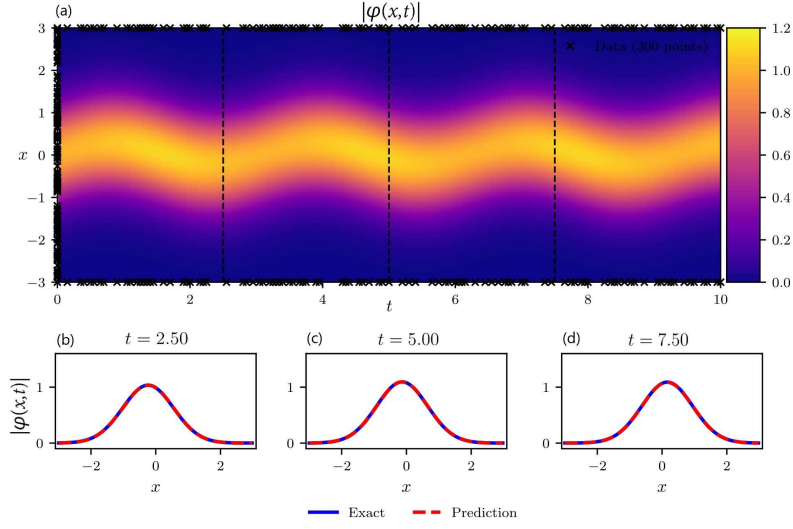}}}
\end{center}
\par
\vspace{-0.15in}
\caption{\protect\small
 The focusing LNLS equation with $\PT$-symmetric potential and initial condition (\ref{aexact}):
(a) The predicted solution $|\varphi(x,t)|$ with $N_I=100$ data points on initial conditions and $N_B=200$ data points on periodic boundary
 conditions (see the cross). We choose $N_S=10,000$ collocation points deduced by Latin Hypercube Sampling strategy.
(b)-(d) Comparisons between the exact and predicted solutions by neural network at $t=2.5,\, 5.0,\, 7.5$. The relative $\mathbb{L}^2-$norm error
of $\varphi(x,t)$ in this case is $4.6204\cdot 10^{-3}$.}
\label{fig3-aexact}
\end{figure}

\begin{figure}[!t]
\begin{center}
\vspace{0.05in} 
{\scalebox{0.56}[0.56]{\includegraphics{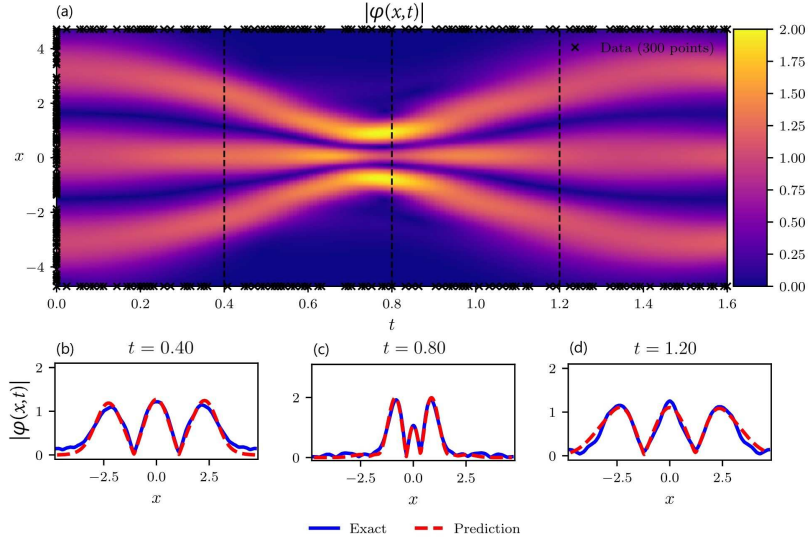}}}
\end{center}
\par
\vspace{-0.15in}
\caption{\protect\small  The focusing LNLS equation with $\PT$-symmetric potential and initial condition (\ref{cos}):
(a) The predicted solution $|\varphi(x,t)|$ with $N_I=100$ data points on initial conditions and $N_B=200$ data points on periodic boundary
 conditions (see the cross). We choose $N_S=10,000$ collocation points deduced by Latin Hypercube Sampling strategy.
(b)-(d) Comparisons between the exact and predicted solutions by neural network at $t=0.4,\,0.8,\, 1.2$. The relative $\mathbb{L}^2-$norm error
of $\varphi(x,t)$ in this case is $1.1713\cdot 10^{-1}$.}
\label{fig4-cos}
\end{figure}

\subsubsection{The periodic initial condition}

In this case, we consider Eq.~(\ref{lnls}) with the periodic initial condition
\bee\label{cos}
 \varphi_0(x)=|\cos(x)|,\quad x\in [-3\pi/2,\, 3\pi/2],
\ene
 the periodic boundary conditions $\varphi(-3\pi/2, t)=\varphi(3\pi/2,t)$, and  parameters  $W_0=0.02$, $V_0=1$, $\sigma=-1$ such that we have $\varphi_{0}(x)=\varphi(x,0)=u_{0}(x)+iv_{0}(x)$ with $u_0(x)=|\cos(x)|$ and $v_0(x)=0$.

We can obtain a training and test data-set by using the above-mentioned Fourier spectral method to simulate Eq.~(\ref{lnls}) with $W_0=0.02,\, V_0=1,\, \sigma=-1$, the initial condition (\ref{cos}), and periodic boundary conditions $\varphi(-3\pi/2,t)=\varphi(3\pi/2,t)$.

 We choose an $11$-layer deep neural network with $50$ neurons per layer and a hyperbolic tangent activation function (\ref{tanh}) to
 jointly represent the latent function $\varphi(x,t)=u(x,t)+iv(x,t)$ where $u(x,t),\, v(x,t)$ represent the real and imaginary parts of $\varphi(x,t)$, respectively. Fig.~\ref{fig4-cos}a displays the magnitude of the predicted solution $|\varphi(x,t)|=\sqrt{u^2(x,t)+v^2(x,t)}$ for the given
locations of the initial-boundary conditions.  The whole training time of this result is $40083$ seconds. The relative $\mathbb{L}^2-$norm error
of $\varphi(x,t)$ in this case is $1.1713\cdot 10^{-1}$. The relative $\mathbb{L}^2-$norm errors  of $u(x,t)$ and $v(x,t)$, respectively, are $1.6250\cdot 10^{-1}$ and $1.6204\cdot 10^{-1}$. Figs.~\ref{fig4-cos}(b-d) exhibit the comparisons between exact and predicted solutions for distinct time instants $t=0.4,\, 0.8,\, 1.2$.

\begin{figure}[!t]
\begin{center}
{\scalebox{0.56}[0.56]{\includegraphics{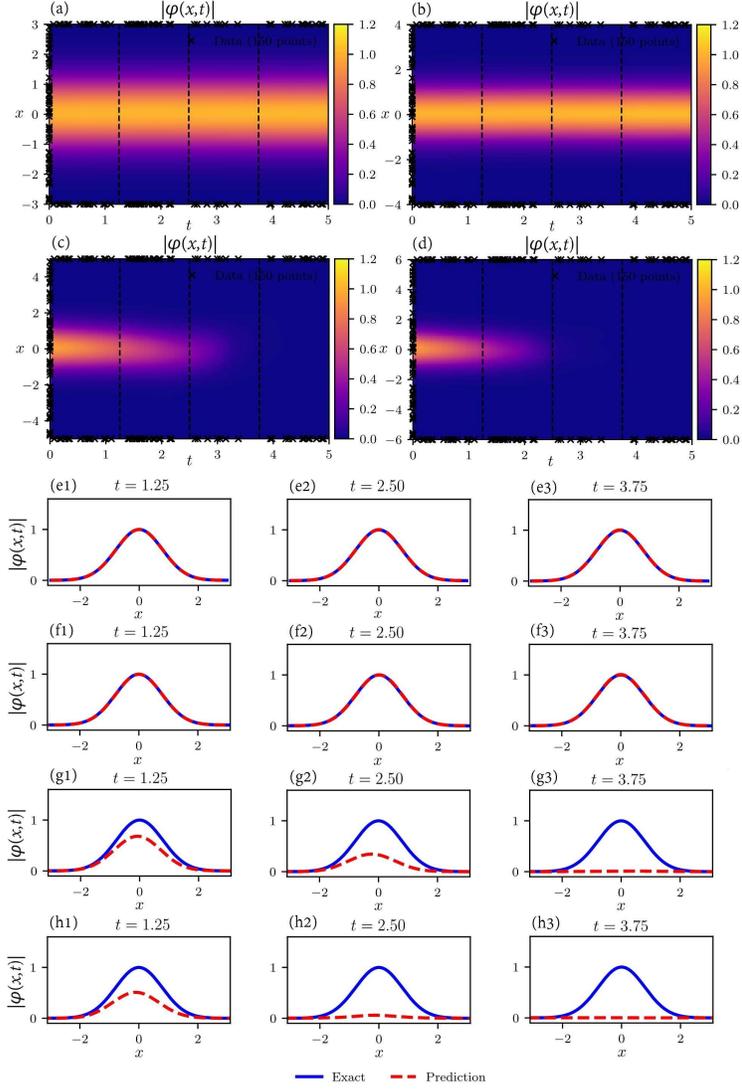}}}
\end{center}
\par
\vspace{-0.15in}
\caption{\protect\small
 The space width influence on the accuracy of deep learning in the LNLS equation with $\PT$-symmetric potential and IBV in Case 2.2.2: (a)-(d) exhibit the deep learning results in four interval widths $L=3,\, 4,\, 5,\, 6$, respectively. We choose randomly $N_I=50$ sampling points on the initial boundary, $N_B=100$ sampling points on the periodic boundary conditions,  and $N_S=8,000$ collocation points deduced by Latin Hypercube Sampling strategy. The comparisons of the exact and predicted solutions for three times $t=1.25,\, 2.5\,,\ 3.75$ in (e1)-(e3): $L=3$; (f1)-(f3): $L=4$; (g1)-(g3): $L=5$; (h1)-(h4): $L=6$.}
\label{fig5-com}
\end{figure}

\section{Comparisons of applications of the PINNs in the LNLS equation}

\subsection{The effects of spatial widths on the learning ability of PINNs}

In what follows, we will discuss the effect of spatial widths on the learning ability of PINNs. In order to simplify this problem, we use the same example in Case 1 of Sec. 3.1. We consider the wave evolution on the distinct intervals including $L=3,\, 4,\, 5, \, 6$, and require the
5000 Adam optimizations and 5000 L-BFGS optimizations.

We use an $11$-layers deep neural network with $50$ neurons per layer and a hyperbolic tangent activation function (\ref{tanh}) to
jointly represent the latent function $\varphi(x,t)$. In the first two rows of Fig.~\ref{fig5-com}, we exhibit the wave propagations for distinct spatial widths and same parameters via the deep learning of PINNs. Obviously, the accuracy of PINNs learning decreases as the space width increases. Their training times are 5001.1s, 4758.5s, 5050.3s, and 4717.5s, respectively.

In the other four rows of Fig.~\ref{fig5-com}, we respectively exhibit the states at three moments for four space widths, which also display that
the predicted solutions are better agree with the exact solutions as the  space width become larger. Table~\ref{table1}
shows all accuracy of the experimental results.

\begin{figure}[!t]
\begin{center}
\vspace{-0.05in} 
{\scalebox{0.56}[0.56]{\includegraphics{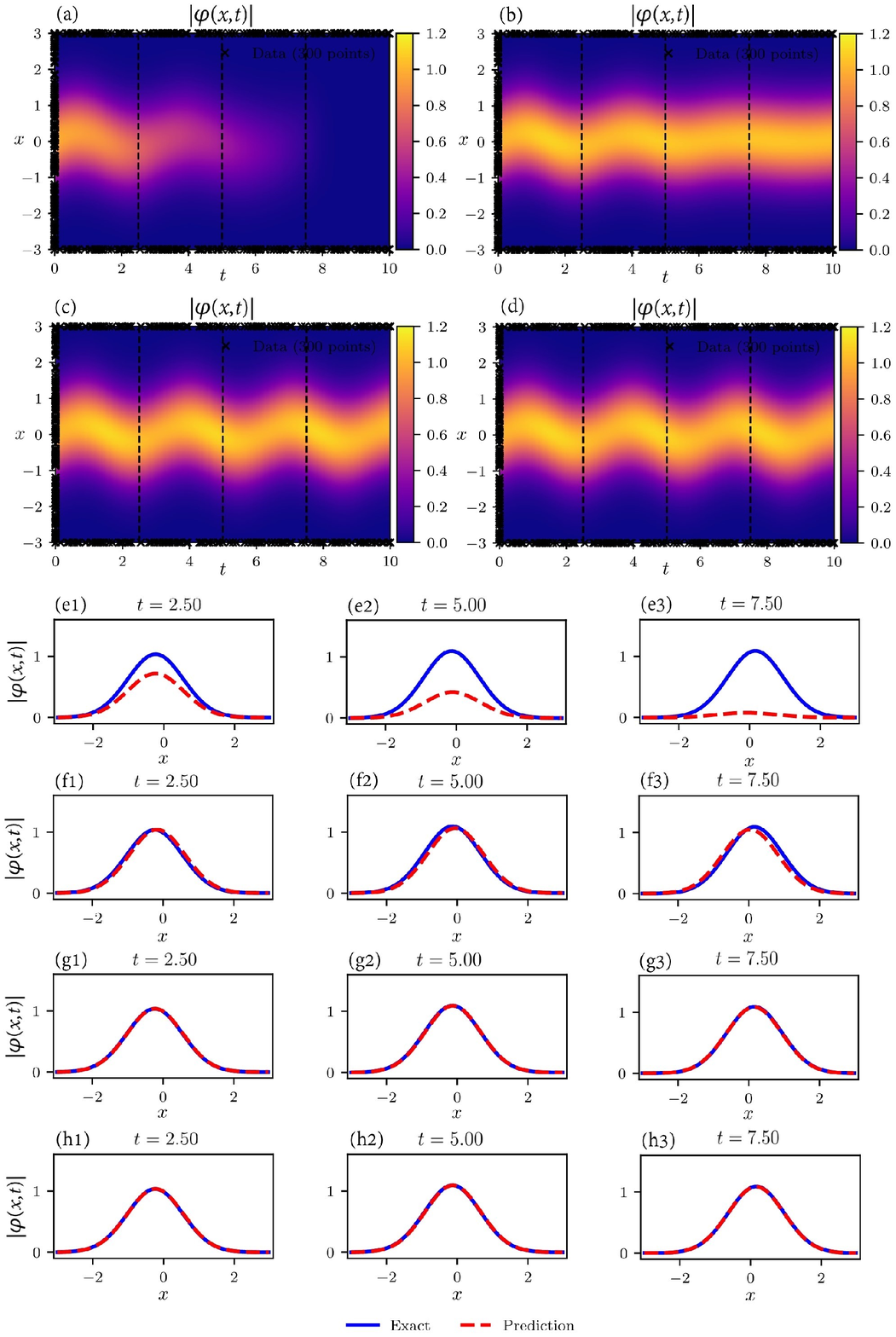}}}
\end{center}
\par
\vspace{-0.15in}
\caption{\protect\small
The optimized step influence on the accuracy of deep learning in the LNLS equation with $\PT$-symmetric potential and IBV in Case 2.2.2:
(a)-(d) The deep learning results in four optimized steps $N=15,000, 30,000, 45,000, 60,000$. We select randomly $N_I=100$ sampling points on initial boundary, $N_B=200$ sampling points on spatial boundary and $N_S=10000$ sampling points in the interior of the region.
The comparisons of the exact and predicted solutions for three times $t=2.50,\, 5.00,\, 7.50$ and (e1)-(e3): $N=15,000$; (f1)-(f3): $N=30,000$; (g1)-(g3): $N=45,000$; (h1)-(h4): $N=60,000$.}
\label{fig6-com}
\end{figure}

\subsection{The effects of optimized steps on the learning ability of PINNs}

In this subsection, we will discuss the effect of optimized steps on PINNs learning ability in the LNLS equation with $\PT$-symmetric potential.
To simplify this problem, we use the same example in (\ref{aexact}). We consider the evolution in different optimized steps: $N=60000, 45000, 30000, 15000$.  We using a 5 hidden layers deep neural network with 90 neurons per layer and a tangent activation function to jointly represent $\varphi(x,t)$. The training steps of these four cases are $60000, 45000, 30000, 15000$ Adam optimizations, and $60000, 45000, 30000, 15000$ L-BFGS optimizations. There training time are 36767s, 32852s, 28295s, and 15512s, respectively (see the first two rows of Fig.~\ref{fig6-com}).  In the other four rows of Fig.~\ref{fig6-com}, we respectively exhibit the states at three moments for four optimized steps, which also display that
the predicted solutions are better agree with the exact solutions as the optimized step become larger. Table~\ref{table2}
shows all accuracy of the experimental results.

\begin{table}[!t]
\centering
\caption{The comparisons between the errors and widths.\vspace{0.05in}}
\begin{tabular}{c|cccc}
\hline
 \diagbox{$\mathbb{L}^2$ error}{Width} &  L=3 & L=4 & L=5 & L=6 \\
    \hline\\
   loss function & 4.34$\times10^{-5}$ & 2.09$\times10^{-4}$ & 8.32$\times10^{-3}$ & 8.69$\times10^{-3}$ \\ \\
   \hline\\
    $u$ & 1.84$\times10^{-2}$ & 2.02$\times10^{-2}$ & 8.79$\times10^{-1}$ & 8.95$\times10^{-1}$ \\ \\
   \hline\\
    $v$ & 1.28$\times10^{-2}$ & 1.19$\times10^{-2}$ & 8.92$\times10^{-1}$ & 9.08$\times10^{-1}$ \\ \\
   \hline\\
   $\varphi$ & 5.64$\times10^{-3}$ & 8.62$\times10^{-3}$ & 7.13$\times10^{-1}$ & 8.03$\times10^{-1}$ \\ \\
   \hline
\end{tabular}
\label{table1}
\end{table}

\begin{table}[!t]
\centering
\caption{The comparisons between the errors and steps.\vspace{0.05in}}
\begin{tabular}{c|cccc}
\hline
 \diagbox{$\mathbb{L}^2$ error}{Step} &  N=15000 & N=30000 & N=45000 & N=60000 \\
    \hline \\
   loss function & 2.76$\times10^{-3}$ & 7.27$\times10^{-5}$ & 5.70$\times10^{-6}$ & 5.16$\times10^{-6}$  \\ \\
   \hline\\
    $u$ & 9.65$\times10^{-1}$ & 1.25$\times10^{-1}$ & 1.47$\times10^{-2}$ & 1.21$\times10^{-2}$ \\ \\
   \hline\\
    $v$ & 9.69$\times10^{-1}$ & 1.29$\times10^{-1}$ & 1.52$\times10^{-2}$ & 1.23$\times10^{-2}$ \\ \\
   \hline\\
    $\varphi$ & 6.77$\times10^{-1}$ & 9.80$\times10^{-2}$ & 1.18$\times10^{-2}$ & 1.04$\times10^{-2}$ \\ \\
   \hline
\end{tabular}
\label{table2}
\end{table}

\section{Data-derived discovery of the LNLS equation with $\PT$-symmetric harmonic potential}

In what follows, we apply the PINNs deep learning method of data-driven discovery of PDEs~\cite{raissi2019} to identify the parameters of the focusing ($\sigma=-1$) LNLS equation with $\PT$-symmetric harmonic potential (\ref{lnls}). The method of data-driven discovery of PDEs is similar to one solving the data-driven solutions of PDEs.

\subsection{Discovering the coefficients of $\PT$-symmetric harmonic potential}

 Here we identify the coefficient of potential $V_{0}, W_{0}$ in the focusing LNLS equation with $\PT$-symmetric harmonic potential
  \begin{eqnarray}\label{lnls-d1}
    i\psi_t=-\psi_{xx}+(V_{0}x^2+iW_{0}x)\psi-\psi\ln(|\psi|^2),
\end{eqnarray}
 where $V_0,\,W_0$ are the unknown real-valued parameters.

  Let $\varphi(x,t)=u(x,t)+iv(x,t)$ with $u(x,t),\,v(x,t)$ being its real and imaginary parts, respectively, and  the PINNs $F(x,t)=iF_u(x,t)-F_v(x,t)$ with $-F_v(x,t),\,F_u(x,t)$ being its real and imaginary parts, respectively, be
 \bee
 \begin{array}{l}
 F(x,t):=i\varphi_t+\varphi_{xx}-(V_0x^2+iW_0x)\varphi+\varphi\ln(|\varphi|^2), \v\\
 F_u(x,t):= u_t + v_{xx} - W_0xu - V_0x^2v + v\ln(u^2 + v^2), \v\\
 F_v(x,t):= v_t - u_{xx} + V_0x^2u - W_0xv - u\ln(u^2 + v^2),
 \end{array}
  \ene
 Then the deep neural network is used to train $\{u(x,t),\, v(x,t)\}$ and parameters $(V_0,\, W_0)$ by minimizing the mean squared error loss
 \bee \label{2-loss}
 {\rm TL}={\rm TL}_{uv}+{\rm TL}_D
 \ene
 with
\begin{equation}
\begin{aligned}
    {\rm TL}_{uv}&=\frac{1}{N_D}\sum_{j=1}^{N_D}\left(|u(x^j,t^j)-u^j|^2+|v(x^j,t^j)-v^j|^2\right),\\
    {\rm TL}_D &=\frac{1}{N_D}\sum_{j=1}^{N_D}\left(|F_u(x^j,t^j)|^2+|F_v(x^j,t^j)|^2\right).
    \end{aligned}
\end{equation}
where $\{x^j,\, t^j,\, u^j,\,v^j\}_{i=1}^{N_D}$ represents the training data on $u(x,t),\, v(x,t)$, and $u(x^j,t^j),\, v(x^j,t^j)$ are real and imaginary parts of the exact solution $\varphi(x,t)=u(x,t)+iv(x,t)$ given by Eq.~(\ref{solu}) with $A=V_0=1,\, W_0=0.2,\, \sigma=-1$ in $(x,t)\in [-3,3]\times[0, 10]$.

To study the data-driven discovery of the LNLS equation (\ref{lnls-d1}) for $V_0,\, W_0$, we generate a  training data-set by randomly choosing
$N_D=10,000$ points in the $(x,t)$ domain,  $[-3,3]\times[0, 10]$ from the exact solution $\psi(x,t)$ for $V_0=1,\, W_0=0.2$. Then the obtained data-set is applied to train an $11$-layer deep NN with $50$ neurons per hidden layer to find the approximate parameters $V_0,\, W_0$ in terms of minimizing the mean squared loss of (\ref{2-loss}). We here use the 20000-step Adam optimization and 20000-step L-BFGS optimization.

For the unknown $V_0,\, W_0$,  Table~\ref{table3} illustrates the correct LNLS equation and identified LNLS equation under the distinct sense of the clean data or the data with $1\%$ noise, and the errors of $V_0,\, W_0$.

\begin{table}[!t]
\centering \caption{Comparison of the correct LNLS equation and the identified ones deduced by deep learning $V_0$ and $W_0$, and their
  errors.\vspace{0.05in}}
\begin{tabular}{c|cccc}
\hline
 \diagbox{\,\,\,\,PDE\,\,\,}{Item} & Equation & Error of $V_{0}$ & Error of $W_{0}$ \\
    \hline \\
  Correct & $i\psi_t=-\psi_{xx}+(x^2+0.2i x)\psi-\psi\ln(|\psi|^2)$ & 0  & 0 \\
  \hline \\
    identified (clean data) & $i\psi_t\!=\!-\psi_{xx}\!+\!(1.00601 x^2\!+\!0.18655ix)\psi\!-\!\psi\ln(|\psi|^2)$ & 6.01$\times10^{-3}$ & 1.34$\times10^{-2}$ \\
    \hline \\
    identified ($1\%$ noise) & $i\psi_t\!=-\!\psi_{xx}\!+\!(0.99249x^2\!+\!0.19985ix)\psi\!-\!\psi\ln(|\psi|^2)$ & 7.51$\times10^{-3}$ & 1.49$\times10^{-4}$ \\
    \hline
\end{tabular}
\label{table3}
\end{table}
\begin{table}[!h]
\centering \caption{Comparison of the correct LNLS equation and the identified ones deduced by deep learning $\alpha$ and $\sigma$, and their
  errors.\vspace{0.05in}}
\begin{tabular}{c|cccc}
\hline
 \diagbox{\,\,\,\, PDE\,\,\,}{Item} & Equation & Error of  $\alpha$ & Error of $\sigma$ \\
    \hline \\
   Correct  & $i\psi_t=-\psi_{xx}+(x^2+0.2i x)\psi-\psi\ln(|\psi|^2)$ & 0 & 0 \\
    \hline \\
    identified (clean data) & $i\psi_t\!=\!-1.00054\psi_{xx}\!+\!(x^2\!+\!0.2i x)\psi\!-\!0.99641\psi\ln(|\psi|^2)$ & 5.43$\times10^{-4}$ & 3.59$\times10^{-3}$ \\
    \hline \\
    identified (1\% noise) & $i\psi_t\!=\!-1.00306\psi_{xx}\!+\!(x^2\!+\!0.2i x)\psi\!-\!0.99003\psi\ln(|\psi|^2)$ & 3.06$\times10^{-3}$ & 9.97$\times10^{-3}$\\
    \hline
\end{tabular}
\label{table4}
\end{table}

 \subsection{Discovering the dispersive and nonlinear coefficients of the LNLS equation}

 In what follows we will identify the dispersive and nonlinear coefficients of the focusing LNLS equation with $\PT$-symmetric harmonic potential in the form
\begin{eqnarray}\label{iden-PDE}
    i\psi_t=-\alpha\psi_{xx}+(x^2+0.2ix)\psi+\sigma\psi\ln(|\psi|^2),
\end{eqnarray}
 where $\alpha,\, \sigma$ are the unknown real-valued parameters.

 Let $\varphi(x,t)=u(x,t)+iv(x,t)$ with $u(x,t),\,v(x,t)$ being its real and imaginary parts, respectively, and  the PINNs $F(x,t)=iF_u(x,t)-F_v(x,t)$ with $-F_v(x,t),\,F_u(x,t)$ being its real and imaginary parts, respectively, be
 \bee
 \begin{array}{l}
 F(x,t):=i\varphi_t+\alpha\varphi_{xx}-[x^2+0.2ix]\varphi-\sigma\varphi\ln(|\varphi|^2), \v\\
 F_u(x,t):= u_t + \alpha v_{xx} - 0.2xu - x^2v -\sigma v\ln(u^2 + v^2), \v\\
 F_v(x,t):= v_t - \alpha u_{xx} + x^2u - 0.2xv +\sigma u\ln(u^2 + v^2),
 \end{array}
  \ene
   Then the deep neural network is used to train $\{u(x,t),\, v(x,t)\}$ and parameters $(\alpha,\, \sigma)$ by minimizing the mean squared error loss given by Eq.~(\ref{2-loss}).

 To study the data-driven discovery of the LNLS equation (\ref{lnls-d1}) for $\alpha,\, \sigma$, we generate a  training data-set by randomly choosing $N_D=10,000$ points in the $(x,t)$ domain,  $[-3,3]\times[0, 10]$ from the exact solution $\psi(x,t)$ for $\alpha=-\sigma=1$. Then the obtained data-set is applied to train an $11$-layer deep NN with $50$ neurons per hidden layer to find the approximate parameters  $\alpha,\, \sigma$ in terms of minimizing the mean squared loss of (\ref{2-loss}). We here also choose the 20000-step Adam optimization and 20000-step L-BFGS optimization.

  For the unknown  $\alpha,\, \sigma$,  Table~\ref{table4} illustrates the correct LNLS equation and identified LNLS equation under the distinct sense of the clean data or the data with $1\%$ noise, and the errors of $\alpha,\, \sigma$.

\section{Conclusions and discussions}

In conclusion, we have used the PINNs deep learning method to solve the LNLS equation with $\PT$-symmetric harmonic potential and distinct
initial-boundary value conditions. We find that the deep learning method is powerful in the LNLS equation by comparing these results and ones deduced from the Fourier spectral method. Moreover, we also verify the PINNs learning ability by solving the LNLS equation with $\PT$-symmetric harmonic potential under the distinct conditions: 1) we fix the optimized step and change the space width; 2) we fix the  space width and change the optimized step. Moreover, we also apply the PINNs deep learning method to effectively investigate the data-driven discovery of the LNLS equation with $\PT$-symmetric harmonic potential. Moreover, the PINNs deep learning method can also be extended to other nonlinear wave equations with $\PT$-symmetric potentials.



\v \noindent {\bf Acknowledgements} \v

This work is partially supported by the NSFC under Grant Nos. 11731014 and 11925108.


\begin{thebibliography}{99}

\bibitem{dl} Y. LeCun, Y. Bengio, and G. Hinton, Deep learning, Nature 521 (2015) 436.

\bibitem{dl2}  I. Goodfellow, Y. Bengio, and A. Courville, {\em Deep learning} (MIT Press, 2016).

\bibitem{ad1} D. Gay, Semiautomatic differentiation for efficient gradient computations. In H. M. Buecker, {et al.} (Eds.), Automatic differentiation: Applications, theory, and implementations (Springer, New York, 50 (2005) 147158).

\bibitem{ad2}  A. G. Baydin, B. A. Pearlmutter, A. A. Radul, and J. M. Siskind, Automatic differentiation in machine learning: a survey,
J. Mach. Learn. Res. 18 (2017) 5595-5637.

\bibitem{ad3} C. C. Margossian, A review of automatic differentiation and its effcient implementation, WIREs Data Mining Knowl. Discov. 9 (2019) e1305.

\bibitem{nn1} M. Dissanayake and N. Phan-Thien, Neural-network-based approximations for solving partial differential equations, Commun. Numer. Methods Eng. 10 (1994) 195-201.

\bibitem{nn2} B. P. van Milligen, V. Tribaldos, and J. Jim\'enez, Neural network differential equation and plasma equilibrium solver, Phys. Rev. Lett. 75 (1995) 3594-3597.

\bibitem{nn3} I. E. Lagaris, A. Likas, and D. I. Fotiadis, Artificial neural networks for solving ordinary and partial differential equations, IEEE Trans. Neural Networks 9 (1998) 987-1000.


\bibitem{siri2018} J. Sirignano and K. Spiliopoulos, DGM: A deep learning algorithm for solving partial differential equations, J. Comput. Phys. 375 (2018) 1339-1364.


\bibitem{raissi2018} M. Raissi, Deep hidden physics models: deep learning of nonlinear partial differential equations, J. Mach. Learn. Res. 19 (2018) 932-955.

\bibitem{raissi2019} M. Raissi, P. Perdikaris, G. E. Karniadakis, Physics-informed neural networks: a deep learning framework for solving forward and inverse problems involving nonlinear partial differential equations, J. Comput. Phys. 378 (2019) 686-707.

\bibitem{pang2019} G. Pang, L. Lu, and G. E. Karniadakis, fPINNs: Fractional physics-informed neural networks, SIAM J. Sci. Comput. 41 (2019) A2603-A2626.

\bibitem{yang2018} L. Yang, D. Zhang, and G. E. Karniadakis, Physics-informed generative adversarial networks for stochastic differential equations, arXiv preprint arXiv:1811.02033 (2018).

\bibitem{zhang2019} D. Zhang, L. Guo, and G. E. Karniadakis, Learning in modal space: Solving time-dependent stochastic PDEs using physics-informed neural networks, arXiv preprint arXiv:1905.01205 (2019).

\bibitem{zhang2019b} D. Zhang, L. Lu, L. Guo, and G. E. Karniadakis, Quantifying total uncertainty in physics- informed neural networks for solving forward and inverse stochastic problems, J. Comput. Phys. 397 (2019) 108850.

\bibitem{nabian2018} M. A. Nabian and H. Meidani, A deep neural network surrogate for high-dimensional random partial differential equations, arXiv preprint arXiv:1806.02957 (2018).

\bibitem{chri2019} C. Beck, W. E, and A. Jentzen, Machine learning approximation algorithms for high-dimensional fully nonlinear partial differential equations and second-order backward stochastic differential equations, J. Nonlinear Sci. 29 (2019) 1563-1619.

\bibitem{lu2019} L. Lu, X. Meng, Z. Mao, G. E. Karniadakis, DeepXDE: A deep learning library for solving differential equations, arXiv:1907.04502.

\bibitem{mich2020} C. Michoski, M. Milosavljevi\'c and T. Oliver et al., Solving differential equations using deep neural networks, Neurocomputing, https://doi.org/10.1016/j.neucom.2020.02.015


\bibitem{rudy2017} S. H. Rudy, S.L. Brunton, J. L. Proctor, J. N. Kutz, Data-driven discovery of partial differential equations, Sci. Adv. 3 (2017) e1602614.

\bibitem{raissi2017} M. Raissi, P. Perdikaris, G. E. Karniadakis, Numerical Gaussian processes for time-dependent and non-linear partial differential
equations, arXiv:1703.10230 (2017).

\bibitem{raissi2017b} M. Raissi, P. Perdikaris, G. E. Karniadakis, Machine learning of linear differential equations using Gaussian processes, J. Comput. Phys. 348 (2017) 683-693.
\bibitem{tensorflow} M. Abadi, P. Barham, J. Chen, {\it et al.,} Tensorflow: A system for large-scale machine learning, in 12th USENIX Symposium on Operating Systems Design and Implementation, 2016, pp. 265-283.

\bibitem{liu89} D. C. Liu, J. Nocedal, On the limited memory BFGS method for large scale optimization, Math. Program. 45 (1989) 503.

\bibitem{back} D. E. Rumelhart, G. E. Hinton, and R. J. Williams, Learning representations by back-propagating errors, Nature 323 (1986) 533-536.

\bibitem{pt1} C. M. Bender, S. Boettcher, Real spectra in non-Hermitian Hamiltonians having $\PT$ symmetry, Phys. Rev. Lett. 80 (1998) 5243.

\bibitem{pt2} C. M. Bender, S. Boettcher, P. N. Meisinger, $\PT$-symmetric quantum mechanics, J. Math. Phys. 40 (1999) 2201.

\bibitem{pt3} Z. Ahmed, Real and complex discrete eigenvalues in an exactly solvable one-dimensional complex $\PT$-invariant potential, Phys. Lett. A 282  (2001) 343-348.

\bibitem{pt4} Z. Musslimani, K. G. Makris, R. El-Ganainy, D. N. Christodoulides, Optical solitons in PT periodic potentials, Phys. Rev. Lett. 100  (2008) 030402.

\bibitem{pt5}  A. Guo, G. Salamo, D. Duchesne, R. Morandotti, M. Volatier-Ravat, V. Aimez, G. Siviloglou, D. Christodoulides, Observation of PT-symmetry breaking in complex optical potentials, Phys. Rev. Lett. 103 (2009) 093902.

\bibitem{pt6}  C. E. R\"uter, K. G. Makris, R. El-Ganainy, D. N. Christodoulides, M. Segev, D. Kip, Observation of parity-time symmetry in optics, Nat. Phys. 6  (2010) 192-195.

\bibitem{pt7}  A. Regensburger, C. Bersch, M.-A. Miri, G. Onishchukov, D. N. Christodoulides, U. Peschel, Parity-time synthetic photonic lattices, Nature 488 (2012) 167-171.

\bibitem{pt8} S. Nixon, L. Ge, J. Yang, Stability analysis for solitons in PT-symmetric optical lattices, Phys. Rev. A 85 (2012) 023822.

\bibitem{miha12} Y. He and D. Mihalache, Spatial solitons in parity-time symmetric mixed linear-nonlinear optical lattices: Recent theoretical results, Rom. Rep. Phys. 64 (2012) 1243.

\bibitem{pt9} Z. Yan, Complex PT-symmetric nonlinear Schr\"odinger equation and Burgers equation, Phil. Tran. R. Soc. A 371 (2013) 20120059.

\bibitem{pt10}  J. Yang, Symmetry breaking of solitons in one-dimensional parity-time-symmetric optical potentials, Opt. Lett. 39  (2014) 5547-5550.

\bibitem{pt11}  Z. Yan, Z. Wen, V. V. Konotop, Solitons in a nonlinear Schr\"odinger equation with PT-symmetric potentials and inhomogeneous nonlinearity: Stability and excitation of nonlinear modes, Phys. Rev. A 92  (2015) 023821.

\bibitem{boris15}  J. D'Ambroise, P. G. Kevrekidis, and B. A. Malomed, Staggered parity-time-symmetric ladders with cubic nonlinearity, Phys.
Rev. E 91 (2015) 033207.

\bibitem{yan16} Z. Yan, Y. Chen, Z. Wen, On stable solitons and interactions of the generalized Gross-Pitaevskii equation with PT- and non-PT-symmetric potentials, Chaos 26 (2016) 083109.

\bibitem{pt12}  V. V. Konotop, J. Yang, D. A. Zezyulin, Nonlinear waves in PT-symmetric systems, Rev. Mod. Phys. 88 (2016) 035002.

\bibitem{pt13} S. V. Suchkov, A. A. Sukhorukov, J. Huang, S. V. Dmitriev, C. Lee, Y. S. Kivshar, Nonlinear switching and solitons in PT-symmetric photonic systems, Laser Photon Rev. 10 (2016) 177-213.

\bibitem{yan17} Z. Yan, Y. Chen, The nonlinear Schr\"odinger equation with generalized nonlinearities and PT-symmetric potentials: Stable solitons, interactions, and excitations, Chaos 27 (2017) 073114.

\bibitem{chen17} Y. Chen, Z. Yan, Stable parity-time-symmetric nonlinear modes and excitations in a derivative nonlinear Schr\"odinger equation, Phys. Rev. E 95, 012205 (2017)

\bibitem{pt14} Y. Chen, Z. Yan, D. Mihalache, B. A. Malomed, Families of stable solitons and excitations in the PT-symmetric nonlinear Schrodinger equations with position-dependent effective masses,
    Scientific Reports 7 (2017) 1257.

\bibitem{abdu18} F. K. Abdullaev and R. M. Galimzyanov, Optical solitons in periodically managed PT-symmetric media, Optik 157 (2018) 353.

\bibitem{chen18} Y. Chen, Z. Yan, Multi-dimensional stable fundamental solitons and excitations in PT-symmetric harmonic-Gaussian potentials with unbounded gain-and-loss distributions, Commun Nonlinear Sci Numer Simulat 57 (2018) 34-46.

\bibitem{pt15} B. Liu, L. Li, D. Mihalache, Effects of periodically-modulated third-order dispersion on periodic solutions of nonlinear Schrodinger equation with complex potential, Rom. Rep. Phys. 70 (2018) 409.

\bibitem{pt16} P. Li, J. Li, B. Han, H. Ma, D. Mihalache, PT-symmetric optical modes and spontaneous symmetry breaking in the space-fractional Schrodinger equation, Rom. Rep. Phys. 71 (2019) 106.

\bibitem{pt17} Y. Chen, Z. Yan, D. Mihalache, Stable flat-top solitons and peakons in the PT-symmetric $\delta$-signum potentials and nonlinear media, Chaos 29 (2019) 083108.

\bibitem{pt18} B. A. Malomed, D. Mihalache, Nonlinear waves in optical and matter-wave media: A topical survey of recent theoretical and experimental results, Rom. J. Phys. 64 (2019) 106.

\bibitem{lse1} I. Bialynicki-Birula and J. Mycielski, Wave equations with logarithmic nonlinearities, Bull. Acad. Polon. Sci. C1. I11 23, 461 (1975).

\bibitem{lse2} I. Bialynicki-Birula and J. Mycielski, Nonlinear wave mechnics, Ann. Phys. (N.Y.) 100 (1976) 62.

\bibitem{lse3} I. Bialynicki-Birula and J. Mycielski,  Gaussons: Solitons of the logarithmic Schr\"odinger equation, Phys. Scr. 20 (1979) 539.

\bibitem{ds} E. S. Hern\'andez, B. Remaud, General properties of gausson-conserving descriptions of quantal damped motion, Physica A 105 (1981) 130.

\bibitem{np} E. F. Hefter, Application of the nonlinear Schr\"odinger equation with a logarithmic inhomogeneous term to nuclear physics, Phys. Rev. A 32 (1985) 1201.

\bibitem{np2} A. Nassar, New method for the solution of the logarithmic nonlinear Schr\"odinger equation via stochastic mechanics, Phys. Rev. A 33 (1986) 3502.

\bibitem{qo} H. Buljan, {\it et al.,} Incoherent white light solitons in logarithmically saturable noninstantaneous nonlinear media, Phys. Rev. E 68 (2003) 036607.

\bibitem{tdp} S. De Martino, M. Falanga, C. Godano, G. Lauro, Logarithmic Schr\"odinger-like equation as a model for magma transport, EPL 63 (2003) 472.

\bibitem{other1} K. G. Zloshchastiev, Logarithmic nonlinearity in theories of quantum gravity: Origin of time and observational consequences, Gravit. Cosmol. 16 (2010) 288.

\bibitem{other4} L. Calaca, A. T. Avelar, D. Bazeia, W. B. Cardoso, Modulation of localized solutions for the Schr\"odinger equation with logarithm nonlinearity, Commun Nonlinear Sci Numer Simulat 19 (2004) 2928-2934.

\bibitem{other2} J. A. Pava and N. Goloshchapova, Stability of standing waves for NLS-log equation with $\delta$-interaction, Nonlinear Differ. Equ. Appl. 24 (2017) 27.

\bibitem{other3} R. Carles, I. Gallagher, Universal dynamics for the defocusing logarithmic Schr\"odinger equation, Duke Math. J. 167 (2018) 1761.

\bibitem{shuai19} W. Shuai, Multiple solutions for logarithmic Schr\"odinger equations, Nonlinearity 32 (2019) 2201.

\bibitem{stein87} M. Stein, Large sample properties of simulations using Latin hypercube sampling, Technometrics 29 (1987) 143-151.

\bibitem{yang2010} J. Yang, {\em  Nonlinear Waves in Integrable and Nonintegrable Systems} (SIAM, Philadelphia, 2010).


\end{thebibliography}
\end{document}